\documentclass[]{aa}
\usepackage{graphicx}
\def\solar {\ifmmode_{\mathord\odot} \else $_{\mathord\odot}$\fi}
\def\Msol {\ifmmode {\,{\it M}\solar} \else $\,M$\solar\fi}
\begin{document}
   \title{LP~349-25: a new tight M8V binary
     \thanks{Based in part on observations made at
      Canada-France-Hawaii Telescope, operated by the National Council
      of Canada, the Centre National de la Recherche Scientifique de
      France and the University of Hawaii, and in part on observations
      at the Paranal observatory, Chile, in ESO program 073.C-0155.}
      }

   \author{T. Forveille\inst{1,2}   
      \and J.-L.~Beuzit\inst{2}
      \and P. Delorme\inst{1,2} 
      \and D. S\'egransan\inst{3}  
      \and X. Delfosse\inst{2}      
      \and G. Chauvin\inst{4} 
      \and T. Fusco\inst{5}
      \and A.-M. Lagrange\inst{2}
      \and M. Mayor\inst{3}     
      \and G. Montagnier\inst{2}
      \and D. Mouillet\inst{6}  
      \and C. Perrier\inst{2}
      \and S. Udry\inst{3} 
      \and J. Charton\inst{2}
      \and P. Gigan\inst{7}
      \and J.-M. Conan\inst{5}
      \and P. Kern\inst{2}
      \and G. Michet\inst{7}
   }

   \offprints{T.~Forveille: \email{Thierry.Forveille@cfht.hawaii.edu}}

   \institute{Canada-France-Hawaii Telescope Corporation, PO Box 1597, 
               Kamuela, HI 96743, USA\\
               \email{Thierry.Forveille@cfht.hawaii.edu}
         \and Laboratoire d'Astrophysique de Grenoble, BP 53X, F-38041 
               Grenoble Cedex, France\\
               \email{Thierry.Forveille@obs.ujf-grenoble.fr, 
                      Jean-Luc.Beuzit@obs.ujf-grenoble.fr,
		      Philippe.Delorme@obs.ujf-grenoble.fr,
                      Xavier.Delfosse@obs.ujf-grenoble.fr,
                      Anne-Marie.Lagrange@obs.ujf-grenoble.fr,
                      Guillaume.Montagnier@obs.ujf-grenoble.fr,
                      Christian.Perrier@obs.ujf-grenoble.fr,
		      Julien.Charton@obs.ujf-grenoble.fr,
		      Pierre.Kern@obs.ujf-grenoble.fr}
         \and Observatoire de Gen\`eve, 51 Chemin des Maillettes, CH-1290, 
               Switzerland \\
               \email{Damien.Segransan@obs.unige.ch, 
                      Michel.Mayor@obs.unige.ch,
                      Stephane.Udry@obs.unige.ch }
         \and European Southern Observatory, Casilla 19001, Santiago 19, 
               Chile \\
               \email{gchauvin@eso.org}
         \and ONERA-DOTA, 92322 Chatillon, France \\
               \email{Thierry.Fusco@onera.fr,
	              Jean-Marc.Conan@onera.fr} 
         \and Laboratoire d'Astrophysique, Observatoire Midi-Pyren\'ees,
              Tarbes, France\\
               \email{mouillet@bagn.obs-mip.fr}
	 \and Laboratoire d'Etudes Spatiales et d'Instrumentation en 
	      Astrophysique, 5 place Jules Janssen, F-92195 Meudon Cedex, 
	      France\\
	       \email{Pierre.Gigan@obspm.fr, 
	        Genevieve.Michet@obspm.fr}
   }       

   \date{Received ; accepted }

   \abstract{We present the discovery of a tight M8V binary, with a
     separation of only 1.2~astronomical units, obtained with the
     PUEO and NACO adaptive optics systems, respectively at the CFHT
     and VLT telescopes. The estimated period of LP~349-25 is
     approximately 5 years, and this makes it an excellent candidate for a
     precise mass measurement.

   }

   \maketitle
%

\section{Introduction}
Thanks to persistent efforts with ground-based adaptive optics 
and spectroscopy (\cite{forveille1999}, \cite{delfosse1999a}, 
\cite{delfosse1999b}, \cite{segransan2000}), as well as with {\it HST} 
(\cite{torres1999}, \cite{benedict2000}; \cite{benedict2001}; 
\cite{hershey1998}), over 30~M~dwarfs now have published masses
with 10\% precision or better. As a result, the empirical Mass-Luminosity
relation is now fairly well constrained down to 0.1~solar mass 
(\cite{delfosse2000}). The near-IR relations for M dwarfs are tight and 
agree very well with theoretical predictions (\cite{baraffe1998}). By contrast,
the $V$~band relation diverges significantly from those models below
$\sim$0.5 solar mass, and has considerable intrinsic dispersion.
The motivation for additional measurements in that mass range is therefore
now shifting towards characterizing -- and understanding -- that
dispersion around the mean relation. 

Below 0.1~solar mass on the other hand, empirical masses are much 
scarcer. Many binaries are now known in that mass range,
and their separations are on average much tighter (typically $<$10~AU)
than in more massive systems (e.g. \cite{close2003}, \cite{bouy2003}). 
Nonetheless, those that are currently known mostly have moderately 
long periods, $\sim$20~years and beyond, which reflect typical distances of 
$\sim$20-30pc and the resolution limit of the observations. To our knowledge,
the only objects with published dynamical masses well below 0.1~solar mass 
are the components of Gl~569BC (\cite{lane2001}, \cite{zapatero2004})
and 2MASSW J0746425+2000321 (\cite{bouy2004}).
Both orbits are still preliminary, with grade 4 in the Washington
Double Star catalog. The observations of 2MASS0746 only cover 35\%
of its period, albeit at a very favourable phase, while Gl~569BC has full
orbital coverage but still somewhat sparse sampling.
Perhaps more importantly, both systems are young enough that at least one
of their components is actually below the Brown Dwarf limit (0.070~solar 
mass, \cite{chabrier2000}), in spite of only moderately late spectral 
types. The age of 2MASS0746 is not independently determined, and the
properties of Gl~569A can only constrain that of Gl~569BC to a broad
interval (\cite{zapatero2004}), over which the model luminosity of
the brown dwarf evolves by an order of magnitude. 
In any comparison with theory, age therefore enters as an unwelcome 
free parameter, and reduces the diagnostic value of those two 
binaries. A few additional systems are being followed, such as LHS~1070
(\cite{leinert1994}; \cite{leinert2001}) and Gl~494  (\cite{beuzit2004}), 
but identifying additional late-M dwarfs binaries with periods under 
$\sim$10 years remains critically important.

Here we present the discovery of one such system, LP~349-25, using the
adaptive optics systems of the CFHT and VLT telescopes. Section~2 
presents the observations and the data analysis, while Section~3 briefly
discusses the properties of the system.

%
%
%
%


\section{Observations and data reduction}
   \begin{figure*}
   \centering
   \begin{tabular}{cc}
     \includegraphics[width=5cm,angle=-90]{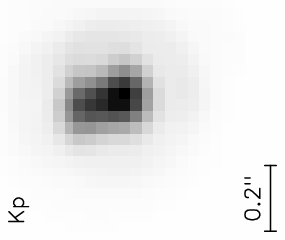} &
     \includegraphics[width=5cm,angle=-90]{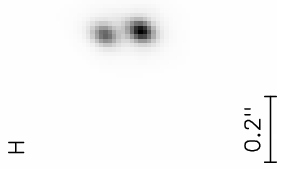} \\
   \end{tabular}
   \caption{Adaptive optics images of LP~349-25 with CFHT through a 
      $K'$ filter (left) and with the VLT through an $H$ filter (right).
      The scale is indicated by a $0.2''$ bar, and North is up and East left. 
      \label{fig_oa}}
    \end{figure*}

\subsection{CFHT observations}
The discovery observations were carried out on July 3$^{rd}$ 2004 at
the 3.6-meter Canada-France-Hawaii Telescope (CFHT), using the CFHT
Adaptive Optics Bonnette (AOB) and the KIR infrared camera. The AOB,
also called PUEO after the sharp-visioned Hawaiian owl, is a
general-purpose adaptive optics (AO) system based on F. Roddier's
curvature concept (\cite{roddier1991}). It is mounted at the telescope
F/8 Cassegrain focus, and cameras or other instruments are then
attached to it (\cite{arsenault1994}; \cite{rigaut1998}). The
atmospheric turbulence is analysed by a 19-element wavefront curvature
sensor and the correction applied by a 19-electrode bimorph mirror.
Modal control and continuous mode gain optimization
(\cite{gendron1994}; \cite{rigaut1994}) maximize the quality of the AO
correction for the current atmospheric turbulence and guide star
magnitude. For our observations a dichroic mirror diverted the visible
light to the wavefront sensor while the KIR science camera
(\cite{doyon1998}, named after a cocktail drink) recorded 
near-infrared light.
The KIR plate scale is 34.85$\pm$0.10 per pixel, for a total field size of 
$36\arcsec \times 36\arcsec$ (KIR on-line users manual). Excellent 
atmospheric conditions prevailed during the observation ($\sim0.55''$ seeing 
in the $V$ band).
The observation sequence consisted of 5 individual 15s exposures at 
each position of an $8''$ square+center offset pattern.
The resulting images are excellent in spite of LP~349-25's faintness 
($V$=17.5, and 40~ADUs/cycle on the wavefront sensor), and its duplicity 
was obvious in real time at the telescope.

\subsection{VLT observations}
Confirmation observations of LP~349-25 were performed on September
26$^{th}$ 2004 with the NACO instrument at VLT UT4 (ESO Very Large
Telescope, Paranal Chile). NACO consists of the NAOS adaptive optics
system (\cite{rousset2003}; \cite{lagrange2003}), providing
diffraction-limited images in the near infrared, and of the CONICA
science camera (\cite{lenzen1998}), equipped with a 1024~$\times$~1024
ALLADIN detector covering the 1-5~$\mu$m spectral domain.  The main
technical features of NAOS are a piezo-stack deformable mirror with
185 actuators and a separate tip-tilt mirror, two selectable
Shack-Hartmann wavefront sensors operating either in the optical
(450-950 nm) or in the near-IR (800-2500 nm) range, both featuring up
to $14 \times 14$ subapertures.  The LP~349-25 observations used the
NAOS IR wavefront sensor, under clear sky and average turbulence
conditions ($0.6''$ seeing and 7 ms coherence time).  They were performed
through the standard $H$ broad-band filter and used the S13 CONICA
camera, which provides a 13.27~mas/pixel sampling (NACO on-line users
manual). The
observation sequence consisted of pairs of ~7s exposures acquired on a
7~positions random offset pattern within a $5''$ jitter box.

\subsection{Data reduction}
The reduction was performed within the ECLIPSE package
(\cite{devillard1997}). The individual raw data were flat-fielded
using a normalised gain map, derived from images of the illuminated
dome at CFHT, and from sky images taken at sunset for the VLT. The sky
signal was estimated from a median across the jittered images. The
individual flat-fielded images were then corrected from the sky image
and stacked using a cross correlation algorithm.

After this cosmetic processing, we used the point-source mode of the
MISTRAL myopic deconvolution package (\cite{mugnier2004}) to extract
the coordinates and intensities of the two stars, from which we
derived the parameters of interest, separation, P.A. and relative
photometry. The astrometric calibration was derived from the standard
Orion field (\cite{mccaughrean1994}) and the HIP~482 wide binary.  
This verified the expected pixel scale of both instruments. 
NACO was, as expected, found oriented within 0.1 degree of North,
and KIR was found rotated by $-2.0\pm$0.2~degrees. 
Centroiding errors and imperfect knowledge
of the point-spread function completely dominate our uncertainty
budget at the small separation of LP~349-25.

Figure~\ref{fig_oa} displays the two reduced images and Table~\ref{tab_oa} 
summarizes the extracted parameters.

\begin{table}
  \caption[]{Adaptive optics measurement of LP~349-25.}
   \label{tab_oa}
   \begin{center}
     \begin{tabular}{|l|c|r|c|c|} \hline
       $\rho$  & $\theta$ & $\Delta$m   & Date & Filt. \\
          $''$     & $\deg$   &             &      &     \\
        \hline 
        0.125$\pm$0.010  & 12.7$\pm$2.0 & 0.26$\pm$0.05 & 03 Jul 2004 & $K'$ \\ 
        0.107$\pm$0.010  &  7.1$\pm$0.5 & 0.38$\pm$0.05 & 26 Sep 2004 & $H$\\ 
        \hline
      \end{tabular}
    \end{center}
\end{table}

\section{Discussion}
LP~349-25 is a recent addition to the solar neighborhood inventory, in
spite of its figuring in the Luyten Two Tenth Catalog
(\cite{luyten1980}).  It was first recognised as a nearby star by
\cite{gizis2000},
during a spectroscopic survey of candidate cool nearby stars selected
from red 2MASS/POSS colours. They derive an M8V spectral type, and estimate
a photometric distance of 8.4~pc from the relatively insensitive $J-K_s$
colour index. \cite{reid2003} use narrow band spectral indices and $J$ band 
photometry to derive a more precise distance, 7.7$\pm$0.8~pc.
We approximately correct that determination for the new companion, and 
adopt a distance for the system of 10.1$\pm$1.2~pc. 

The companion is bright in the infrared, $K'$=10.46, and on that ground 
alone it is highly improbable that it is a background star. At the position of 
LP~349-25 the density of the 2MASS catalog for $K<$11 is 60~sources/square 
degree. The probability to find such a bright star within even an arcsecond 
of LP~349-25 is therefore only $1.5~10^{-6}$. Additionally, the companion 
cannot be much bluer than the primary, or the system would 
have produced a stronger signal on the visible wavefront sensor. 
Galactic reddening behind LP~349-25 ($l_{II}=115.81\deg$, 
$b_{II}=-40.22\deg$) is very small (approximately $E(B-V)$=0.06, 
\cite{burstein82}), and a background star would thus need to be
intrisically as red as LP~349-25. This would make it either an 
unrelated red dwarf in the immediate solar neighbourhood, or 
a halo giant at $\sim$20~kpc. Both possibilities are highly unlikely.

The proper motion and parallax are large 
($+399.8\pm$5.5~mas/yr and $-177.2\pm$5.5~mas/yr \cite{salim2003}); 
130$\pm$13mas, assuming for the sake of this particular argument that 
the star is actually single), but largely cancel out between 
the dates of our two observations, with a total motion of only $-19\pm$7 
and $-47\pm$4 mas.
The separation of the two components actually
changed by $-14\pm$15~mas and $-16\pm$15~mas.
This is compatible with the expected orbital motion of $\sim$30mas 
(uncertain by a factor of a few), but only helps excluding a background
object at the 2~$\sigma$ level.

LP~349-25 however has been previously examined for multiplicity, with the 
HOKUPAA adaptive optics system on GEMINI (\cite{close2002}), and
it was then found unresolved. The two components were most likely less 
separated at that time (September 18th and 19th 2001), or perhaps for this 
particular target \cite{close2002} did not obtain as good an adaptive optics 
correction as we have. Had the system however been significantly wider than
found here, \cite{close2002} would have been able to resolve
it even with degraded correction. Their negative result 
demonstrates that LP~349-25 is not a long period system, which we would
have serendipitously observed close to periastron. It also definitely
ensures that the companion is not a background star, which on that
date would have been separated by $1.3''$, and very obviously resolved.
A background star would 
in addition have been separated by $21.9''$ at the 1954 epoch of 
the blue plate of the first Palomar Survey, and again it would be very easily 
seen.

For late-M and early-L dwarfs, one spectral subtype corresponds
to approximately 0.35~magnitude at $H$ band (e.g. \cite{vrba2004}). The 
observed contrast therefore indicates that the make-up of the pair is
M7.5V+M8.5V or M8V+M9V.
At the 10~pc distance of the system, its $0.12''$ separation translates
to 1.2~AU. If the stars have reached the main sequence (age $>\sim$~1Gyr), 
as implicitly assumed to evaluate the distance, both masses are approximately 
0.08~{\Msol} (\cite{baraffe1998}), just above the Brown Dwarf limit. 
Adopting the main sequence masses, and correcting for the 1.35 statistical 
factor between instantaneous projected separation and semi-major axis 
(\cite{duquennoy1991}), the orbital period is approximately 5~years. 
This estimate obviously has significant uncertainties, but it makes 
LP~349-25 one of the best
candidates for an accurate mass determination below 0.1~\Msol.  We
plan to monitor its relative motion with adaptive optics and will
attempt to obtain a spectroscopic orbit, but a precise trigonometric
parallax and an astrometric orbit will be equally important for the
mass determination.



\begin{thebibliography}{}
  \bibitem[Arsenault et al. 1994]{arsenault1994} Arsenault R., Salmon D., 
    Kerr J., et al., 1994, in SPIE Proceedings 2201, Adaptive
    Optics in Astronomy, ed. M.A. Ealey, F. Merkle, 833 

  \bibitem[Baraffe et al. 1998]{baraffe1998} Baraffe I., Chabrier G.,
    Allard F., Hauschildt P.H., 1998, A\&A 337, 403

  \bibitem[Benedict et al. 2000]{benedict2000} Benedict, G.~F., 
    McArthur, B.~E., Franz, O.~G., Wasserman, L.~H., \& Henry, T.~J.\ 2000, 
    \aj, 120, 1106 

  \bibitem[Benedict et al. 2001]{benedict2001} Benedict, G.~F.,  McArthur, 
    B.~E., Franz, O.~G. et al.\ 2001, \aj, 121, 1607 

  \bibitem[Beuzit et al. 2004]{beuzit2004} Beuzit, J.-L., S\'egransan, D.,
    Forveille, T., et al.\ 2004, \aap, 425, 997 

  \bibitem[Bouy et al. 2003]{bouy2003} Bouy, H., Brandner, W., 
    Mart{\'{\i}}n, E.~L., et al.\ 2003, \aj, 
    126, 1526 

  \bibitem[Bouy et al. 2004]{bouy2004} Bouy, H., Duch\^ene, G., K\"ohler, R.,
    et al.\ 2004, \aap, 423, 341 

\bibitem[Burstein \& Heiles 1982]{burstein82} Burstein, D., \& 
    Heiles, C.\ 1982, \aj, 87, 1165 

  \bibitem[Chabrier et al. 2000]{chabrier2000} Chabrier, G., Baraffe,
    I., Allard, F., \& Hauschildt, P.\ 2000, \apj, 542, 464

  \bibitem[Close et al. 2002]{close2002} Close, L.~M., Siegler, 
   N., Potter, D., Brandner, W., \& Liebert, J.\ 2002, \apjl, 567, L53 

  \bibitem[Close et al. 2003]{close2003} 
    Close, L.~M., Siegler, N., Freed, M., \& Biller, B.\ 2003, \apj, 587, 407 

 \bibitem[Delfosse et al. 1999a]{delfosse1999a} Delfosse, X., 
    Forveille, T., Mayor, M., Burnet, M., \& Perrier, C.\ 1999a, 
    \aap, 341, L63 

  \bibitem[Delfosse et al. 1999b]{delfosse1999b} Delfosse, X., 
    Forveille, T., Udry, S., et al.\ 1999b, 
    \aap, 350, L39 

  \bibitem[Delfosse et al. 2000]{delfosse2000} Delfosse, X., 
    Forveille, T., S{\' e}gransan, D., et al.\ 2000, \aap, 364, 217 
  
  \bibitem[Devillard 1997]{devillard1997} Devillard, 
    N.\ 1997, The Messenger, 87, 5 

  \bibitem[Doyon et al. 1998]{doyon1998} Doyon R., Nadeau D., Vall\'ee
    P., et al.\ 1998, in SPIE Proceedings 3354, Infrared
    Astronomical Instrumentation, ed. A.M. Fowler, 760

  \bibitem[Duquennoy \& Mayor 1991]{duquennoy1991} Duquennoy A., Mayor M., 
    1991, \aap, 248, 485

  \bibitem[Forveille et al. 1999]{forveille1999} Forveille, T., Beuzit, J.-L.,
    Delfosse, X., et al.\ 1999, \aap, 351, 619 

  \bibitem[Gendron \& L\'ena 1994]{gendron1994} Gendron E., L\'ena P., 
    1994, \aap, 291, 337

  \bibitem[Gizis et al. 2000]{gizis2000} Gizis, J.~E., Monet, 
    D.~G., Reid, I.~N., et al.\ 2000, \aj, 120, 1085 

  \bibitem[Hershey \& Taff 1998]{hershey1998} Hershey, J.~L.~\& 
    Taff, L.~G.\ 1998, \aj, 116, 1440 

  \bibitem[Lagrange et al. 2003]{lagrange2003} Lagrange, A.-M., Chauvin, G.,
    Fusco, T., et al.\ 2003, \procspie, 4841, 860 

  \bibitem[Lane et al. 2001]{lane2001} Lane, B.~F., Zapatero 
    Osorio, M.~R., Britton, M.~C., Mart{\'{\i}}n, E.~L., \& Kulkarni, S.~R.\ 
    2001, \apj, 560, 390 

  \bibitem[Leinert et al. 1994]{leinert1994} Leinert, C., Weitzel, 
    N., Richichi, A., Eckart, A., \& Tacconi-Garman, L.~E.\ 1994, \aap, 291, 
    L47 

  \bibitem[Leinert et al. 2001]{leinert2001} Leinert, C., 
    Jahrei{\ss}, H., Woitas, J., et al.\ 2001, \aap, 367, 183 

  \bibitem[Lenzen et al. 1998]{lenzen1998} Lenzen, R., Hofmann, R.,
    Bizenberger, P., \& Tusche, A.\ 1998, \procspie, 3354, 606

  \bibitem[Luyten 1980]{luyten1980} Luyten, W.~J.\ 1980, University 
    of Minnesota Minneapolis, 55, 1 

  \bibitem[McCaughrean \& Stauffer 1994]{mccaughrean1994} McCaughrean, 
    M.~J.~\& Stauffer, J.~R.\ 1994, \aj, 108, 1382 

  \bibitem[Mugnier, Fusco \& Conan 2004]{mugnier2004} Mugnier L.~M., 
    Fusco T. \& Conan J.-M., 2004, J. Opt. Soc. Am. A., 21(10), 1841
                                                                                
  \bibitem[Reid et al. 2003]{reid2003} Reid, I.~N., Cruz, K., Allen, P.,
    et al.\ 2003, \aj, 126, 3007 

  \bibitem[Rigaut et al. 1994]{rigaut1994} Rigaut F., Arsenault R.,
     Kerr J., et al.\ 1994, in
     SPIE Proceedings 2201, Adaptive Optics in Astronomy,
     ed. M.A. Ealey, F. Merkle, 149

  \bibitem[Rigaut et al. 1998]{rigaut1998} Rigaut F., Salmon D., 
    Arsenault R. et al., 1998, \pasp, 110, 152

  \bibitem[Roddier et al. 1991]{roddier1991} Roddier F., Graves J.E., 
    Mc~Kenna D., Northcott M.J., 1991, in SPIE Proceedings 1524, 248

  \bibitem[Rousset et al. 2003]{rousset2003} Rousset, G., Lacombe, F.,
    Puget, P., et al.\ 2003, \procspie, 4839, 140 

  \bibitem[Salim \& Gould 2003]{salim2003} Salim, S.~\& Gould, A.\ 
    2003, \apj, 582, 1011 

  \bibitem[S{\' e}gransan et al. 2000]{segransan2000} S{\' e}gransan, 
    D., Delfosse, X., Forveille, T., et al.\ 2000, \aap, 364, 665 

  \bibitem[Torres et al. 1999]{torres1999} 
    Torres, G., Henry, T.~J., Franz, O.~G., \& Wasserman, L.~H.\ 1999, \aj, 
    117, 562 

  \bibitem[Vrba et al. 2004]{vrba2004} Vrba, F.~J., Henden, A.~A.,
   Luginbuhl, C.~B., et al.\ 2004, 
    \aj, 127, 2948 

  \bibitem[Zapatero Osorio et al. 2004]{zapatero2004} 
    Zapatero Osorio, M.R., Lane, B.L., Pavlenko, Ya., et al.\
    2004, astro-ph/0407334

\end{thebibliography}
\end{document}